\documentclass[12pt ]{article}

\renewcommand{\theequation}
{\arabic{section}.\arabic{equation}}

\makeatletter
\def\eqnarray{ \stepcounter{equation} \let\@currentlabel=\theequation
 \global\@eqnswtrue
 \global\@eqcnt\z@
 \tabskip\@centering
 \let\\=\@eqncr
 $$\halign to \displaywidth\bgroup\@eqnsel\hskip\@centering
 $\displaystyle\tabskip\z@{##}$&\global\@eqcnt\@ne
 \hfil$\displaystyle{{}##{}}$\hfil
 &\global\@eqcnt\tw@$\displaystyle\tabskip\z@{##}$\hfil
 \tabskip\@centering&\llap{##}\tabskip\z@\cr}
\makeatother

\makeatletter
\def\@arrayacol{\edef\@preamble{\@preamble \hskip .5\arraycolsep}}
\def\array{\let\@acol\@arrayacol \let\@classz\@arrayclassz
\let\@classiv\@arrayclassiv \let\\\@arraycr\def\@halignto{}\@tabarray}
\makeatother

\makeatletter
\newcounter{subeqncnt}
\def\thesubeqncnt{\alph{subeqncnt}}
\def\subequations{\begingroup%
   \stepcounter{equation}\edef\@tempa{\theequation}%
   \let\c@equation\c@subeqncnt\c@subeqncnt\z@
   \edef\theequation{\@tempa\noexpand\thesubeqncnt}}

\makeatother

\newcommand{\be}{\begin{equation}}
\newcommand{\ee}{\end{equation}}

\newcommand{\beqa}{\begin{eqnarray}}
\newcommand{\eeqa}{\end{eqnarray}}
\newcommand{\nn}{\nonumber}

\newcommand{\eqref}[1]{(\ref{#1})}

\begin{document}

\setlength{\baselineskip}{7mm}
\begin{titlepage}
\begin{flushright}

{\tt NRCPS-HE-27-2020} \\

\end{flushright}

\begin{center}
{\Large ~\\{\it  Parton Distribution Functions \\
and \\
Tensorgluons 
\vspace{1cm}

}

}

\vspace{1cm}

{\sl Roland Kirschner${}^{a,1}$  and
 George Savvidy${}^{b,2}$

\bigskip
${}^{a}$ \sl Institut f\"ur Theoretische Physik, Universit\"at Leipzig \\ Augustusplatz 10, D-04109 Leipzig, Germany\\
${}^{b}$ \sl Institute of Nuclear and Particle Physics\\ Demokritos National Research Center, Ag. Paraskevi,  Athens, Greece\\

\bigskip

\centerline{$^{1}$~{\footnotesize\it kirschne@itp.uni-leipzig.de},
$^{2}$~{\footnotesize\it savvidy@inp.demokritos.gr}}

}
 
\end{center}
\vspace{30pt}

\centerline{{\bf Abstract}}
We further consider a possibility that inside the proton and, more generally, inside the hadrons, there are additional partons - tensorgluons, which can carry a part of the proton momentum.  The tensorgluons have zero electric charge, like gluons, but have a larger spin. Inside the proton
a nonzero density of the tensorgluons can be generated by the emission of tensorgluons
by gluons. The last mechanism is typical for non-Abelian tensor gauge
theories. The process of gluon splitting suggests that part of the proton momentum
that was carried by neutral partons is shared between vector and tensorgluons.
We derive the regularised evolution equations for the parton distribution functions that take into account these new processes.  In particular, this will allow to solve numerically the extended DGLAP equations and to find out the ratio of densities between gluons and tensorgluons. 
\vspace{12pt}

\noindent

\end{titlepage}



\pagestyle{plain}

\section{\it Introduction}

It was predicted that the Bjorken scaling should be broken by logarithms
of a transverse momentum $Q^2$ and that these
deviations from the scaling law can be computed for the deep inelastic
structure functions \cite{Gross:1973id,Politzer:1973fx,Altarelli:1977zs,Dokshitzer:1977sg,Gribov:1972ri,Gribov:1972rt,
Lipatov:1974qm,Fadin:1975cb,Kuraev:1977fs,Balitsky:1978ic,Savvidy:2013zwa,Cabibbo:1978ez,Gribov:1981ac}.
In the leading logarithmic approximation the results
can be formulated in the parton language \cite{Bjorken:1969ja} by assigning the well determined $Q^2$
dependence to the parton densities \cite{Gross:1973id,Politzer:1973fx,Gross:1973ju,Gross:1974cs,Altarelli:1977zs}.

In this article we shall  further consider a possibility that inside the proton and, more generally,
inside the hadrons there are
additional partons - tensorgluons, which can carry a part of the proton momentum \cite{Savvidy:2014hha,Savvidy:2005fi,Savvidy:2005zm,Savvidy:2005ki,Savvidy:2010vb}.  It was proposed  in \cite{Savvidy:2013zwa} that a possible emission of tensorgluons
inside proton will produce a tensorgluon "cloud"  of neutral  tensorgluon partons
in addition to the quark and gluon "clouds''. Tensorgluons have
zero electric charge, like gluons, but have a larger spin. Inside the proton
a nonzero density of the tensorgluons can be generated by the emission of tensorgluons
by gluons \cite{Savvidy:2005fi,Savvidy:2005zm,Savvidy:2005ki,Savvidy:2010vb}.
The last mechanism is typical for non-Abelian tensor gauge
theories, in which there exists a gluon-tensor-tensor vertex of order $g$.
The number of gluons changes not only
because a quark may radiate a gluon or because a gluon may split into a quark-antiquark
pair or into two gluons \cite{Gross:1973ju,Gross:1974cs,Altarelli:1977zs},
but also because a gluon can split into two tensorgluons
\cite{Savvidy:2005fi,Savvidy:2005zm,Savvidy:2005ki,Savvidy:2010vb,Georgiou:2010mf,Antoniadis:2011rr,Kirschner:2017vqm}.
The process of gluon splitting into tensorgluons suggests that part of the proton momentum
which was carried by neutral partons is shared between vector and tensorgluons.

The proposed model was formulated in terms of a field theory Lagrangian,
which describes the interaction of the gluons with their massless partners of higher
spin, the  tensorgluons \cite{Savvidy:2005fi,Savvidy:2005zm,Savvidy:2005ki,Savvidy:2010vb}.  
The gauge invariant Lagrangian for the lower-rank tensor
gauge fields has the form \cite{Savvidy:2005fi,Savvidy:2005zm,Savvidy:2005ki}:
\beqa\label{totalactiontwo}
{{\cal L}}  =&-&{1\over 4}G^{a}_{\mu\nu }G^{a}_{\mu\nu }
-{1\over 4}G^{a}_{\mu\nu,\lambda}G^{a}_{\mu\nu,\lambda}
-{1\over 4}G^{a}_{\mu\nu}G^{a}_{\mu\nu,\lambda\lambda}\\
&+&{1\over 4}G^{a}_{\mu\nu,\lambda}G^{a}_{\mu\lambda,\nu}
+{1\over 4}G^{a}_{\mu\nu,\nu}G^{a}_{\mu\lambda,\lambda}
+{1\over 2}G^{a}_{\mu\nu}G^{a}_{\mu\lambda,\nu\lambda}+...\nn
\eeqa
where the the field strength tensors have the form:
\beqa\label{fieldstrengthparticular}
G^{a}_{\mu\nu} &=&
\partial_{\mu} A^{a}_{\nu} - \partial_{\nu} A^{a}_{\mu} +
g f^{abc}~A^{b}_{\mu}~A^{c}_{\nu},\\
G^{a}_{\mu\nu,\lambda} &=&
\partial_{\mu} A^{a}_{\nu\lambda} - \partial_{\nu} A^{a}_{\mu\lambda} +
g f^{abc}(~A^{b}_{\mu}~A^{c}_{\nu\lambda} + A^{b}_{\mu\lambda}~A^{c}_{\nu} ~),\nn\\
G^{a}_{\mu\nu,\lambda\rho} &=&
\partial_{\mu} A^{a}_{\nu\lambda\rho} - \partial_{\nu} A^{a}_{\mu\lambda\rho} +
g f^{abc}(~A^{b}_{\mu}~A^{c}_{\nu\lambda\rho} +
 A^{b}_{\mu\lambda}~A^{c}_{\nu\rho}+A^{b}_{\mu\rho}~A^{c}_{\nu\lambda}
 + A^{b}_{\mu\lambda\rho}~A^{c}_{\nu} ~),\nn\\
 ......&.&............................................\nn
\eeqa
The first term in (\ref{totalactiontwo}) corresponds to the standard Yang-Mills Lagrangian.  The expression for the full Lagrangian which describes the infinite tour of interacting high-rank tensor gauge bosons can be found in \cite{Savvidy:2005fi,Savvidy:2005zm,Savvidy:2005ki}. For illustration we shall present the next term of the Lagrangian that describes the interaction of the rank-3 tensor gauge boson 
\beqa\label{thirdranktensorlagrangian}
{{\cal L}} 
=&-&{1\over 4}G^{a}_{\mu\nu,\lambda\rho}G^{a}_{\mu\nu,\lambda\rho}
-{1\over 8}G^{a}_{\mu\nu ,\lambda\lambda}G^{a}_{\mu\nu ,\rho\rho}
-{1\over 2}G^{a}_{\mu\nu,\lambda}  G^{a}_{\mu\nu ,\lambda \rho\rho}
-{1\over 8}G^{a}_{\mu\nu}  G^{a}_{\mu\nu ,\lambda \lambda\rho\rho}+ \nn\\
&+&{1\over 3}G^{a}_{\mu\nu,\lambda\rho}G^{a}_{\mu\lambda,\nu\rho}+
{1\over 3} G^{a}_{\mu\nu,\nu\lambda}G^{a}_{\mu\rho,\rho\lambda}+
{1\over 3}G^{a}_{\mu\nu,\nu\lambda}G^{a}_{\mu\lambda,\rho\rho}+\\
&+&{1\over 3}G^{a}_{\mu\nu,\lambda}G^{a}_{\mu\lambda,\nu\rho\rho}
+{2\over 3}G^{a}_{\mu\nu,\lambda}G^{a}_{\mu\rho,\nu\lambda\rho}
+{1\over 3}G^{a}_{\mu\nu,\nu}G^{a}_{\mu\lambda,\lambda\rho\rho}
+{1\over 3}G^{a}_{\mu\nu}G^{a}_{\mu\lambda,\nu\lambda\rho\rho}.\nn
\eeqa
The characteristic property of the model is that all interaction vertices
between gluons and tensorgluons have the universal {\it dimensionless coupling constant $g$} in four-dimensional space-time (\ref{totalactiontwo}).
That is, the cubic interaction vertices have only first order derivatives and the quartic vertices
have no derivatives at all. These are familiar properties of the standard Yang-Mills field theory.
In order to understand the physical properties of the
model it was important to study the tree-level scattering amplitudes.
A very powerful spinor helicity technique \cite{Berends:1981rb,Kleiss:1985yh,Xu:1986xb,
Gunion:1985vca,Dixon:1996wi,Parke:1986gb,Berends:1987me,
Witten:2003nn,Cachazo:2004by,Cachazo:2004dr,Britto:2004ap,Britto:2005fq,
Benincasa:2007xk,Cachazo:2004kj,Georgiou:2004by,
ArkaniHamed:2008yf,Berends:1988zn,Mangano:1987kp}
was used to calculate  high-order tree-level diagrams with the participation of tensorgluons
in \cite{Georgiou:2010mf}.

Here we shall further develop the regularisation technique proposed earlier for the splitting amplitudes.  The present paper is organised as follows. In section two the basic formulae for
scattering amplitude and splitting functions are recalled, definitions  and notations
are specified, the details of the regularisation scheme are presented. 
In section three we derive the regularised evolution equations for the parton distribution functions that 
take into account the creation of tensorgluons.  Section four contains concluding remarks and summarises the physical consequences of the tensorgluons hypothesis.

\section{\it Splitting Functions}
It was proposed  in \cite{Savvidy:2013zwa} that a possible emission of tensorgluons
inside proton will produce a tensorgluon "cloud"  of neutral  tensorgluon partons
in addition to the quark and gluon "clouds''.
Our goal is to specify the regularisation of the generalised 
DGLAP equations \cite{Altarelli:1977zs,Gribov:1972ri,Gribov:1972rt,Lipatov:1974qm,Fadin:1975cb,Kuraev:1977fs,Balitsky:1978ic,Dokshitzer:1977sg} that was introduced in \cite{Savvidy:2013zwa}.

In the generalized Yang-Mills theory \cite{Savvidy:2005fi,Savvidy:2005zm,Savvidy:2005ki,Savvidy:2010vb} all interaction vertices between gluons and tensorgluons have {\it dimensionless coupling constant}.
Using these vertices  one can compute the scattering amplitudes of gluons and  tensorgluons. The color-ordered
scattering amplitudes involving two tensorgluons of helicities $h =\pm s$,
one negative helicity gluon
and $(n-3)$ gluons of positive helicity were found  in  \cite{Georgiou:2010mf}. 
These scattering amplitudes have been used to extract splitting amplitudes
of gluons and tensorgluons \cite{Antoniadis:2011rr}.  
Since the collinear limits of the scattering amplitudes
are responsible for parton evolution, one can extract from these expressions the
splitting probabilities for  tensorgluons \cite{Savvidy:2013zwa}:
\beqa\label{setoftensorgluon}
P_{TG}(z) &=&  C_2(G)\left[ {z^{2s+1} \over (1-z)^{2s-1}_{+}}  
+{(1-z)^{2s+1} \over z^{2s-1}}   \right],\nn\\
P_{GT}(z) &=&  C_2(G)\left[ {1\over z(1-z)^{2s-1}_{+}} 
+{(1-z)^{2s+1} \over z }   \right],\\
P_{TT}(z) &=&  C_2(G)\left[ {z^{2s+1} \over  (1-z)_{+}} 
+{1 \over (1-z)_{+} z^{2s-1}}  \right]. \nn
\eeqa
The invariant operator $C_2$ for the representation R is defined by the equation
$ t^a t^a  = C_2(R)~ 1 $ and $tr(t^a t^b) = T(R) \delta^{ab}$. These
functions satisfy the relations
\be
P_{TG}(z)=P_{TG}(1-z),~~~P_{GT}(z)=P_{TT}(1-z),~~~~~~~z < 1.
\ee
The kernel $P_{TG}(z)$ has a meaning of variation per unit time of the probability
density of finding a tensorgluon inside the gluon, $P_{GT}(z)$ - of finding gluon inside
the tensorgluon and $P_{TT}(z)$ - of finding tensorgluon inside the tensorgluon.
One should define the regularisation procedure for the singular factors
$(1 - z)^{-2s+1}$ and $ z^{-2s+1}$  reinterpreting them as the  distributions $(1 - z)^{-2s+1}_{+}$ and
$z^{-2s+1}_{+}$. The regularisation has been defined in the following
way  \cite{Savvidy:2013zwa}:
\beqa\label{definition}
\int_{0}^{1} dz {f(z)\over (1 - z)^{2s-1}_+}&=&
\int_{0}^{1} dz {f(z)- \sum^{2s-2}_{k=0} {(-1)^k \over k!} f^{(k)}(1) (1-z)^k \over (1 - z)^{2s-1}},\nn\\
\nn\\
\int_{0}^{1} dz {f(z)\over z ^{2s-1}_+}&=&
\int_{0}^{1} dz {f(z)- \sum^{2s-2}_{k=0} {1 \over k!} f^{(k)}(0) z^k \over z^{2s-1}},\\
\nn\\
\int_{0}^{1} dz {f(z)\over z_+ (1-z)_+}&=&
\int_{0}^{1} dz {f(z)-  (1-z)f(0) - z f(1) \over z  (1-z) },\nn
\eeqa
where $f(z)$ is any test function that is sufficiently regular at the end points
and, as one can see, the defined substraction guarantees the convergence of the integrals.
Using the same arguments as in the standard case \cite{Altarelli:1977zs} we should add the delta function
terms into the definition of the diagonal kernels so that they will completely determine
the behaviour of $P_{qq}(z)$ , $P_{GG}(z)$ and $P_{TT}(z)$ functions.

One should add $\delta (z-1)$  to the diagonal  kernels $P_{qq}(z)$, $P_{GG}(z)$ and  $P_{TT}(z)$
with the coefficients that have been determined by using the momentum sum rule:
\beqa
P_{qq}(z)&=& C_2(R)\left[ {1+z^2 \over (1-z)_+ } + {3\over 2}\delta (z-1)\right],\nn\\
P_{GG}(z) &=&  2 C_2(G)\left[{z \over (1-z)_+}+ {1-z  \over z }+ z(1-z) \right]
+  {\sum_s(12s^2 -1)C_2(G) - 4 n_f T(R)  \over 6}    ~ \delta (z-1),\nn\\
P_{TT}(z) &=&  C_2(G)\left[{z^{2s+1} \over (1- z)_+ }   + {1\over (1-z)_{+} z^{2s-1}}
+ \sum^{2s+1}_{j=1} {1 \over j} ~ \delta (z-1) \right].
\eeqa
Thus we completed the definition of the kernels appearing in the
evolution equations (\ref{evolutionequation}).
For completeness we shall present also quark and gluon splitting functions \cite{Altarelli:1977zs}:
\beqa\label{setofquarkgluon}
P_{Gq}(z) &=& C_2(R){1+(1-z)^2 \over z },\\
P_{qG}(z) &=& T(R)[z^2 +(1-z)^2], 
\eeqa
where $C_2(G)= N, C_2(R)={N^2-1  \over  2 N},  T(R) = {1  \over  2}$ for the SU(N) groups.

At the end of this section we shall discuss, what type of processes could also be included into
the evolution equations (\ref{evolutionequation}). In (\ref{evolutionequation})
we ignore contribution of
the high-spin fermions $\tilde{q}^i$ of spin $s + 1/2$, which are the partners of the
standard quarks \cite{Savvidy:2005fi,Savvidy:2005zm,Savvidy:2005ki,Savvidy:2010vb},
supposing  that they are even heavier than the top quark. That is,
all kernels $P_{q\tilde{q}},P_{G\tilde{q}}, P_{T\tilde{q}},P_{\tilde{q}\tilde{q}},P_{\tilde{q}q},
P_{\tilde{q}G},P_{\tilde{q}T}$ and $P_{Tq}$ with the emission of $\tilde{q}^i$ are taken to be
zero. These terms can be included in the case of very high energies, but at modern energies
it seems safe to ignore these contributions. In the evolution equation
for tensorgluons in (\ref{evolutionequation}) one could also include the
kernels $P_{TT^{'}}$ that describe the emission of  tensorgluons by tensorgluons,
the $TT^{'}T^{''}$ vertex   \cite{Savvidy:2005fi,Savvidy:2005zm,Savvidy:2005ki,Savvidy:2010vb}.
That also can be done, and the number of evolution equations in that case will
tend to infinity. In this article we shall limit ourselves by considering only emissions
that always involve
the standard gluons and spin-2 tensors ignoring infinite "stairs" of transitions
between tensorgluons.

\section{\it Regularisation of Generalised  DGLAP Equations}

The deep inelastic structure functions
can be  expressed in terms of parton densities \cite{Altarelli:1977zs,Gribov:1972ri,Gribov:1972rt,Lipatov:1974qm,Fadin:1975cb,Kuraev:1977fs,Balitsky:1978ic,Dokshitzer:1977sg}.
If $q^i(x, Q^2)$ is the density of quarks of type i (summed over
colors) inside a nucleon target with fraction x of the proton longitudinal momentum
in the infinite momentum frame, then the unpolarised structure functions can
be represented in the following form:
$$
2F_1(x,Q^2)= F_2(x,Q^2)/x= \sum_i e^2_i [q^i(x,Q^2)+\bar{q}^i(x,Q^2)].
$$
The $Q^2$
dependence of the parton densities is described by 
the integro-differential equations for quark  $q^i(x,t)$ and gluon densities  $G(x,t)$,
where $t=\ln(Q^2/Q^2_0)$ \cite{Altarelli:1977zs,Gribov:1972ri,Gribov:1972rt,
Lipatov:1974qm,Fadin:1975cb,Kuraev:1977fs,
Balitsky:1978ic,Dokshitzer:1977sg}. If there
is  an additional emission of tensorgluons in the proton, then one should introduce
the corresponding density $T(x, t)$ of tensorgluons  and 
the integro-differential equations that describe the $Q^2$ dependence
of parton densities in this general case has the following form  \cite{Savvidy:2013zwa}:
\beqa\label{evolutionequation}
{d q^i(x,t)\over dt} &=& {\alpha(t) \over 2 \pi} \int^{1}_{x} {dy \over y}[\sum^{2 n_f}_{j=1} q^j(y,t)~
P_{q^i q^j}({x \over y})+ G(y,t)~ P_{q^i G}({x \over y})] ,\\
{d G(x,t)\over dt} &=& {\alpha(t) \over 2 \pi} \int^{1}_{x} {dy \over y}[\sum^{2 n_f}_{j=1} q^j(y,t)~
P_{G q^j}({x \over y})+ G(y,t) ~P_{G G}({x \over y})+ T(y,t) ~P_{G T}({x \over y}) ],\nn\\
{d T(x,t)\over dt} &=& {\alpha(t) \over 2 \pi} \int^{1}_{x} {dy \over y}[
G(y,t)~ P_{T G}({x \over y}) +  T(y,t)~ P_{T T}({x \over y})].\nn
\eeqa
The $\alpha(t)$ is the running coupling ($\alpha = g^2/4\pi$) and has the following form \cite{Savvidy:2014hha}:
\be\label{strongcouplingcons}
{\alpha \over \alpha(t)} = 1 +b ~\alpha ~t~~,
\ee
where
\be\label{fullbeta}
b =  {\sum_s (-1)^{2s}(12s^2 -1) C_2(G) - 4 n_f T(R) \over 12 \pi} ~
\ee
is the one-loop Callan-Symanzik  coefficient. In particular, the presence of the 
spin-two tensorgluons in the proton will give
\be\label{fullbeta2}
b=  {58 C_2(G) - 4 n_f T(R) \over 12 \pi}.
\ee
The density of tensorgluons $T(x,t)$  changes when
a gluon splits into two tensorgluons or when a tensorgluon radiates a gluon. This
evolution is described by the last equation (\ref{evolutionequation}).

The tensorgluon kernels (\ref{setoftensorgluon})
 are singular at the boundary values similar to the
case of the standard kernels (\ref{setofquarkgluon}),
though  the singularities are of higher order compared to the standard case.
The '+' prescription in 
\beqa\label{evolutionmartix}
P(z)= \left(\begin{array}{ccc}
  P_{qq}(z)&2 n_f P_{qG}(z)&0\\
  P_{Gq}(z)&P_{GG}(z)&P_{GT}(z)\\
 0&P_{TG}(z)&P_{TT}(z)
\end{array} \right)
\eeqa
is defined as 
\beqa
[g(x)]_+=g(x) - \delta(1-x) \int^{1}_{0} g(z)dz,
\eeqa
and so  
\be\label{reggluon}
\int^{1}_{x} f(z) [g(z)]_+ dz = \int^{1}_{x}[f(z)-f(1)]g(z)dz - f(1)\int^{1}_{x}g(z)dz.
\ee
Considering the splitting probabilities for spin two tensorgluons we have to
define '+++' prescription as 
\beqa
[g(x)]_{+++}=g(x) - \delta(1-x) \int^{1}_{0} g(z)dz 
- \delta^{'}(1-x) \int^{1}_{0} g(z)(1-z)dz- \nn\\
- {1\over 2}\delta^{''}(1-x) \int^{1}_{0} g(z)(1-z)^2dz,
\eeqa
and so  
\beqa\label{prescription}
\int^{1}_{x} f(z) [g(z)]_{+++} dz &=& 
\int^{1}_{x}[f(z)-f(1) +f^{'}(1)(1-z)   -{1\over 2}f^{''}(1)(1-z)^2]g(z)dz -\nn\\
&&- \int^{x}_{0}[f(1)-f^{'}(1)(1-z) +{1\over 2}f^{''}(1)(1-z)^2]g(z)dz.
\eeqa
For the helicity-2 tensorgluons, the  $s=2$ in (\ref{setoftensorgluon}), we will have 
\beqa\label{setoftensorgluon2}
P_{TG}(z) &=&  C_2(G)\left[ {z^{5} \over (1-z)^{3}}  
+{(1-z)^{5} \over z^{3}}   \right],\nn\\
P_{GT}(z) &=&  C_2(G)\left[ {1\over z(1-z)^{3}} 
+{(1-z)^{5} \over z }   \right],\\
P_{TT}(z) &=&  C_2(G)\left[ {z^{5} \over  (1-z)} 
+{1 \over (1-z) z^{3}} + \sum^{5}_{j=1} {1 \over j} ~ \delta (1-z)  \right]\nn
\eeqa
and  the regularisation of these kernels can be performed using the regularisation prescription (\ref{prescription}). The regular splitting functions will take the following form:
\beqa\label{reg1}
 &&\int^{1}_{x}P_{TG}(z) f({x\over z})dz = 
\int^{1}_{x}\left[{z^{5} \over (1-z)^{3}}_{+++}  + {(1-z)^{5} \over z^{3}} \right] f({x\over z})dz = I_1+I_2+I_3 ,\\
&&I_1=\int^{1}_{x} {dz \over (1-z)^3}    
\Big( z^5 f({x\over z}) -f(x) +( 5f(x) -xf^{'}(x) ) (1-z)   - \nn\\
&&~~~~~~~~~~~~~~~~~~~~~~~~~~~~~~~~~~~~~~~~~~~~~~~~~~~~~~~~~~ -(10f(x) -4xf^{'}(x)   + {1\over 2}x^2 f^{''}(x) ) (1-z)^2 \Big)  , \nn\\
&& I_2=- \int^{x}_{0}  {dz \over (1-z)^3} \Big(f(x)-(5f(x)-xf^{'}(x) ) (1-z) + \nn\\
&&~~~~~~~~~~~~~~~~~~~~~~~~~~~~~~~~~~~~~~~~~~~~~~~~~~~~~~~~~~+(10f(x) -4xf^{'}(x) 
+{1\over 2}x^2f^{''}(x)) (1-z)^2 \Big) ,  \nn\\
&&I_3= \int^{1}_{x}   {(1-z)^{5} \over z^{3}}  f({x\over z})dz . \nn
\eeqa
For the gluon-tensor splitting function we will get:
\beqa\label{reg2}
 &&  \int^{1}_{x} P_{GT}(z)  f({x\over z})dz =    \int^{1}_{x}\left[   {3-3z +z^2 \over  (1-z)^{3}}_{+++} 
+{1+(1-z)^{5} \over z }             \right] f({x\over z})dz    = J_1+J_2 +J_3,\nn\\
&& J_1=\int^{1}_{x} {dz \over (1-z)^3}
\Big( (3-3z +z^2) f({x\over z}) -f(x) - (f(x) +xf^{'}(x)) (1-z)   -\nn\\
&&~~~~~~~~~~~~~~~~~~~~~~~~~~~~~~~~~~~~~~~~~~~~~~~~~~-(f(x) +2xf^{'}(x)   + {1\over 2}x^2 f^{''}(x)) (1-z)^2 \Big)  , \nn\\
&& J_2=- \int^{x}_{0}  {dz\over (1-z)^3} [f(x) + (f(x)+ x f^{'}(x)(1-z) +(f(x) +2xf^{'}(x) +{1\over 2}x^2f^{''}(x)(1-z)^2] ,\nn\\
&&J_3=\int^{1}_{x}   {1+(1-z)^{5} \over z}  f({x\over z})dz. 
\eeqa
Using the regularisation (\ref{reggluon}) for the tensor-tensor splitting function we will get the following expression:
\beqa\label{reg3}
 &  \int^{1}_{x} P_{TT}(z)  f({x\over z})dz =    \int^{1}_{x}\left[      {1 +z^5 \over  (1-z)}_{+} 
+{1+ z +z^{2} \over z^3 }       + \sum^{5}_{j=1} {1 \over j} ~ \delta (1-z)      \right] f({x\over z})dz = 
\nn  \\
&=\int^{1}_{x} {dz\over (1-z)} \left[ (1 +z^5) f({x\over z}) -2 f(x)  \right]   
      + f(x)\left[ 2 \ln(1-x)  +  {137 \over 60}\right] +\nn\\
&+\int^{1}_{x}   {1+z + z^{2} \over z^3}  f({x\over z})dz .
\eeqa
All new splitting functions which have been added to the standard evolution equation are now well defined and can be calculated using the above equations (\ref{reg1})-(\ref{reg3}).

\section{\it Conclusion}

Let us summarise the physical consequences of the tensorgluons hypothesis.
Among all parton distributions, the gluon density $G(x,t)$ is one of the least
constrained functions since it does not couple directly to the
photon in deep-inelastic scattering measurements of the proton $F_2$ structure function.
Therefore it is only indirectly constrained by scaling violations and by the momentum sum rule
which resulted in the fact that only half of the proton momentum is carried by charged
constituents - the quarks - and that the other part is ascribed to the
neutral constituents. As it was suggested in \cite{Savvidy:2013zwa}, the process of gluon splitting leads to the emission of tensorgluons and therefore a part of the proton momentum that is carried  by the neutral constituents here is shared between gluons and tensorgluons.  The density of neutral partons in the proton is therefore given by the sum of two functions: $G(x,t)+T(x,t)$, where $T(x,t)$ is the density of the tensorgluons.   Thus tensorgluons can carry a part of nucleon momentum together with gluons. Because tensorgluons have a larger spin, they can influence the spin structure of the nucleon. This contribution appears in the next to leading order, compared to the gluons and can provide a substantial screening effect due to the larger spin of tensorgluons. The details can be found in \cite{Konitopoulos:2016xuj}. To disentangle these contributions and to decide which piece of the neutral partons
is the contribution of gluons and which one is of the tensorgluons one should measure the helicities of the neutral components, which seems to be a difficult task. The other test of the proposed model will be
the consistency of the mild $Q^2$ behaviour of the moments of the structure functions with the experimental data.

In supersymmetric extensions of the Standard Model \cite{Fayet:1976et,Fayet:1977yc}
the gluons and quarks have natural partners - gluinos of spin s=1/2 and squarks of spin s=0.
If the gluinos appear as elementary constituents of the hadrons, then the theory predicts the
existence of new hadronic states, the R-hadrons \cite{Farrar:1978xj,Fayet:1978qc}.
These new hadronic states can be produced in ordinary hadronic collisions, and they decay
into ordinary hadrons with the radiation of massless photino - the massless partner of
the photon, which takes out a conserved R-parity. Depending on the model  the
gluinos may be massless or
massive, depending on the remaining unbroken symmetries
and the representation content of the theory. The existing experimental data
give evidence that most probably they have to be very heavy
\cite{Farrar:1978rk,Antoniadis:1982qw}.

It seems that phenomenological limitation on the existence of the R-hadrons
is much stronger that the limitation on the existence of the tensorgluons
inside the ordinary hadrons. This is because gluinos change the statistics of
the ordinary hadrons: the proton has to have a partner, the R-proton, which is a boson
and is indeed a new hadron. 

The existence of tensorgluon partons inside the proton
does not predict a new hadronic state, a proton remains a proton. The tensorgluons
will alternate the parton distribution functions of a proton. The question is to which extent the tensorgluons will change the parton distribution functions. The regularisation of the splitting amplitudes developed above (\ref{reg1})-(\ref{reg3}) will allow a numerical solution of  the generalised DGLAP evolution equations (\ref{evolutionequation}) for the parton distribution functions that takes into account the processes of emission of tensorgluons by gluons. The integration can be performed using the algorithms developed in \cite{Botje:2010ay,Bertone:2017kne,Miyama:1995bd,Vogt:2004ns}  and will allow to find out the ratio of densities between gluons and tensorgluons.

\end{document}